# Improving the Performance of Multi-class Intrusion Detection Systems using Feature Reduction

3 AUTHORS, INCLUDING:

Yasmen Essam
Suez Canal University

**1** PUBLICATION   **0** CITATIONS

SEE PROFILE

# Improving the Performance of Multi-class Intrusion Detection Systems using Feature Reduction

Yasmen Wahba[1], Ehab ElSalamouny [2] and Ghada ElTaweel [3]

[1] Faculty of Computers and Informatics, Suez Canal University
Ismailia, Egypt

[2] Faculty of Computers and Informatics, Suez Canal University
Ismailia, Egypt

[3] Faculty of Computers and Informatics, Suez Canal University
Ismailia, Egypt

**Abstract**
Intrusion detection systems (IDS) are widely studied by researchers nowadays due to the dramatic growth in network-based technologies. Policy violations and unauthorized access is in turn increasing which makes intrusion detection systems of great importance. Existing approaches to improve intrusion detection systems focus on feature selection or reduction since some features are irrelevant or redundant which when removed improve the accuracy as well as the learning time. In this paper we propose a hybrid feature selection method using Correlation-based Feature Selection and Information Gain. In our work we apply adaptive boosting using naïve Bayes as the weak (base) classifier. The key point in our research is that we are able to improve the detection accuracy with a reduced number of features while precisely determining the attack. Experimental results showed that our proposed method achieved high accuracy compared to methods using only 5-class problem. Correlation is done using Greedy search strategy and naïve Bayes as the classifier on the reduced NSL-KDD dataset.

## 1. Introduction

Securing networks from intrusions or attacks is becoming harder as the network technologies are rapidly growing. The number of distributed denial of service (DDoS) attacks has increased by 90 percent as reported by the state of the internet security 2014 report, while the average attack duration increased by 28 percent [1]. Organizations often deploy a firewall as a first line of defense in order to protect their private network from malicious attacks, but there are several ways to bypass the firewall which makes Intrusion detection system a second line of defense and a way to monitor the network traffic for any possible threat or illegal action [2].

Intrusion detection systems generally fall into two main types: anomaly detection systems and Misuse detection systems [3]. In anomaly based techniques, the classification is based on rules where any attempt that falls out of the normal behaviour is treated as an attack, unlike misuse detection techniques where a list of signatures of known attacks is kept in the system and compared with captured data, which implies that novel attacks cannot be detected.

Extensive attention is given to examining several ways for improving the performance of IDS and feature selection methods proved to be an effective way for enhancing the performance by reducing the feature set and removing irrelevant and redundant features. Feature selection is a crucial step in most classification problems which reduces the learning time and enhances the predictive accuracy [4].

Feature selection algorithms are classified into wrapper and filter methods. While wrappers usually provide the best feature set and deliver high accuracy, they are computationally expensive since they repeatedly invoke a predetermined induction algorithm. It thus becomes unpractical to apply wrappers when having a large dataset. Filter methods are more preferred in the sense that they do not involve any learning algorithm which makes them much faster compared to wrappers.

In this research paper we use the simple correlation based feature selection (CFS) which is a filter method that selects the best feature subset according to some evaluation function where features are assumed to be conditionally independent. Based on the former assumption, CFS is not guaranteed to select all relevant features when there are strong feature dependencies[5]. So we used Information Gain (IG) as a ranking step for the rest of the features that were not selected by CFS in the first phase. Then the final feature set is the total of features selected from CFS and those who were ranked high by the Information gain measure based on a predetermined threshold. Finally, classification is done using the method of adaptive boosting which is applied to naïve Bayes as a base classifier.

We conducted several experiments to verify the effectiveness of our feature selection methods using 23 classes, results were compared to approaches using only 5 classes. Experimental results showed that our proposed method performs well in terms of detection rate as well as keeping a low false positive rate whilst using the full set of attacks. Also we showed that using adaptive boosting with naïve Bayes classifier greatly improves the learning process and enhances the detection rates for almost all of the attacks.

Previous works are studied in the next section. Information Gain is described in Section 3. Section 4 explains the CFS feature selection method. Afterwards, in Section 5 discretization is discussed as a preprocessing step. Adaptive boosting is presented in Section 6. The NSL-KDD dataset which we use in our experiments is presented in Section 7. Experiments and results are given in Section 8. Finally, conclusions and future works are presented in Section 9.

## 2. Related Works

Data mining approaches are being widely studied with intrusion detection systems as a way to identify hidden and interesting patterns in network traffic data [6]. Techniques like classification, clustering and regression have been used to build intrusion detection systems. Several clustering-based techniques have been studied for the design of IDS, [7] used multiple centroid-based clustering algorithms to identify new attack instances. Efforts for using neural networks in the IDS field showed promising results as reported by the SANS Institute Reading Room [8]. J. Ryan, M. Lin [9] presented a new way of applying neural network believing that the attacker leaves a print each time he uses the network.

For the sake of improving the performance of classification, adaptive boosting was used by many researchers. [10][11] used naïve Bayes as a weak learner enhanced with AdaBoost and achieved extremely low False Positive rate. Research showed that using an ensemble of classification techniques usually deliver better results than individual approaches. [12] suggested using an ensemble of Artificial Neural Networks (ANNs), Support Vector Machines (SVMs) and Multivariate Adaptive Regression Splines (MARS). Whereas [13] proposed a hybrid approach using a Radial Basis Function (RBF) and Support Vector Machine (SVM).

Due to high dimensionality of network data, feature selection techniques gained a huge attention as a pre-processing phase prior to classification. [14] have used correlation to eliminate redundant records and then fed the reduced dataset to a 3-layer neural network. Authors in [15] showed how feature reduction can improve the detection accuracy, they reduced the features using information gain, gain ratio and correlation. Experiments showed that combining feature selection methods could possibly improve classification accuracy, [4] suggested using a hybrid feature selection using information gain and symmetrical uncertainty, while [16] used feature Quantile filter and Chi-Squared to reduce the number of features. Others introduce Genetic Algorithms along with Linear Discriminant Analysis as a hybrid feature selection method [17].

Authors in [18] proposed a sequential search strategy for feature selection through determining the importance of a given attribute by simply removing it and recording the performance, if performance increased then the feature is unimportant and thus shall be removed. Since One technique may give good results for one dataset while under-perform for another, TOPSIS [19] was suggested to rank various feature selection techniques based on a confidence value between 0 and 1, the higher the confidence value means a more preferred technique.

## 3. Information Gain

Information gain is used as a measure for evaluating the worth of an attribute based on the concept of entropy, the higher the entropy the more the information content. Entropy can be viewed as a measure of uncertainty of the system [5]. The Entropy of a discrete feature Y is defined as

$$H(Y) = -\sum_{y \in Y} p(y) \log_2 (p(y))$$

(1)

Information gain for two attributes X and Y is defined as

$$IG(X,Y) = H(Y) - H(Y|X) = H(X) - H(X|Y)$$

(2)

As clear from the above equation, Information gain is a symmetrical measure—that is, the amount of information gained about Y after observing X is equal to the amount of information gained about X after observing Y.

In our proposed method we evaluate the information gain between individual features and the class. Accordingly, features are ranked by their relevancy to the class. The higher the gain, the more relevant the feature for determining the class labels.

Information gain is also widely used in classification using decision trees to decide the ordering of attributes in the decision tree. The feature with the highest information gain is considered as more discriminative than other features and is placed at the root of the tree.

## 4. Correlation based feature selection (CFS)

CFS is considered as one of the simplest yet effective feature selection methods. It is based on the assumption that features are conditionally independent given the class, where feature subsets are evaluated based on the following hypothesis[5]:
A good feature subset is one that contains features highly correlated with (predictive of) the class, yet uncorrelated with (not predictive of) each other.
One of the advantages of CFS is that it is a filter algorithm, which makes it much faster compared to a wrapper selection method since it does not need to invoke the learning algorithm [4].
The Evaluation function is described by the following equation

$$M_s = \frac{k\overline{r_{cf}}}{\sqrt{k + k(k-1)\overline{r_{ff}}}} \qquad (3)$$

Where $M_s$ is the heuristic "merit" of a feature subset S containing K features, $\overline{r_{cf}}$ is the mean feature-class correlation, and $\overline{r_{ff}}$ is the average feature-feature intercorrelation.
Experiments showed that CFS not only runs faster compared to the wrapper but also produce comparable results, and might outperform the wrapper on small datasets. However, when features are highly dependent on each other, CFS can fail to select all the relevant features [1].

## 5. Discretization

Discretization is the process of quantizing Continuous attributes by grouping those values into a number of discrete intervals [20]. Some classifiers only deal with discrete data, and thus discretization becomes a crucial step before classification. Discretization can be classified into supervised and unsupervised methods. In this paper we choose the popular method of Entropy Minimization Discretization (EMD) introduced by Fayyad and Irani [21]. We also remark that EMD is the default method that is used in the Weka tool [22].
Researchers showed that discretization greatly improves the overall performance of classification as well as saving storage space since the discretized data require less space [23].

## 6. AdaBoost

The AdaBoost algorithm was first introduced by Freund and Schapire [24]."Boosting" is a general method used to improve the performance of any learning algorithm. The main idea of boosting lies in calling the base algorithm repeatedly where in each round incorrectly classified examples are assigned higher weights so that the algorithm focus on the hard examples in the successive rounds [25].
Adaptive boosting is used in conjunction with one or more weak learners in order to enhance their performance.

## 7. Intrusion data set

Our dataset is the NSL-KDD (http://iscx.ca/NSL-KDD/) which is suggested to solve some of the problems in the original KDD99 dataset[26]. One of the most important deficiencies in the KDD data set is the huge number of redundant records, which causes the learning algorithms to be biased towards the frequent records. The dataset contains 41 features which are listed in the table below

Table 1: List of features in NSL-KDD dataset

| No. | Feature name | Type |
|---|---|---|
| 1 | Duration | Continuous |
| 2 | Protocol-type | Discrete |
| 3 | Service | Discrete |

| 4 | Flag | Discrete |
|---|---|---|
| 5 | Src-bytes | Continuous |
| 6 | Dst-bytes | Continuous |
| 7 | Land | Discrete |
| 8 | Wrong-fragment | Continuous |
| 9 | Urgent | Continuous |
| 10 | Hot | Continuous |
| 11 | Num-failed-logins | Continuous |
| 12 | Logged-in | Discrete |
| 13 | Num-compromised | Continuous |
| 14 | Root-shell | Continuous |
| 15 | Su-attempted | Continuous |
| 16 | Num-root | Continuous |
| 17 | Num-file-creations | Continuous |
| 18 | Num-shells | Continuous |
| 19 | Num-access-files | Continuous |
| 20 | Num-outbound-cmds | Continuous |
| 21 | Is-host-login | Discrete |
| 22 | Is-guest-login | Discrete |
| 23 | Count | Continuous |
| 24 | Srv-count | Continuous |
| 25 | Serror-rate | Continuous |
| 26 | Srv-serror-rate | Continuous |
| 27 | Rerror-rate | Continuous |
| 28 | Srv-rerror-rate | Continuous |
| 29 | Same-srv-rate | Continuous |
| 30 | Diff-srv-rate | Continuous |
| 31 | Srv-diff-host-rate | Continuous |
| 32 | Dst-host-count | Continuous |
| 33 | Dst-host-srv-count | Continuous |
| 34 | Dst-host-same-srv-rate | Continuous |
| 35 | Dst-host-diff-srv-rate | Continuous |
| 36 | Dst-host-same-src-port-rate | Continuous |
| 37 | Dst-host-srv-diff-host-rate | Continuous |
| 38 | Dst-host-serror-rate | Continuous |
| 39 | Dst-host-srv-serror-rate | Continuous |
| 40 | Dst-host-rerror-rate | Continuous |
| 41 | Dst-host-srv-rerror-rate | Continuous |

In our work we extracted only 62984 records, where 53% are normal records and the 47% are distributed among the different attack types. These attacks fall into the following four main categories [27].

7.1. Denial of service (Dos), where attempts are to suspend services of a network resource making it unavailable to its intended users by overloading the server with too many requests to be handled.

7.2. Probe attacks, where the hacker scans the network with the aim of exploiting a known vulnerability.

7.3. Remote-to-Local (R2L) attacks, where an attacker tries to gain local access to unauthorized information through sending packets to the victim machine.

7.4. User-to-Root (U2R) attacks, where an attacker gains root access to the system using his normal user account to exploit vulnerabilities.

Table 2 shows the distribution of the attacks in our dataset, given 62984 instances.

Table 2: Attacks distribution and corresponding class

| *Attack* | *No. of records* | Class |
|---|---|---|
| land | 6 | Dos |
| neptune | 20750 | Dos |
| smurf | 1327 | Dos |
| pod | 87 | Dos |
| back | 502 | Dos |
| teardrop | 437 | Dos |
| portsweep | 1489 | Probe |
| ipsweep | 1814 | Probe |
| satan | 1829 | Probe |
| nmap | 743 | Probe |
| multihop | 5 | R2L |
| spy | 1 | R2L |
| phf | 3 | R2L |
| warezclient | 469 | R2L |
| guess_passwd | 27 | R2L |
| ftp_write | 4 | R2L |
| warezmaster | 13 | R2L |
| imap | 6 | R2L |
| buffer_overflow | 17 | U2R |
| loadmodule | 3 | U2R |
| perl | 1 | U2R |
| rootkit | 7 | U2R |

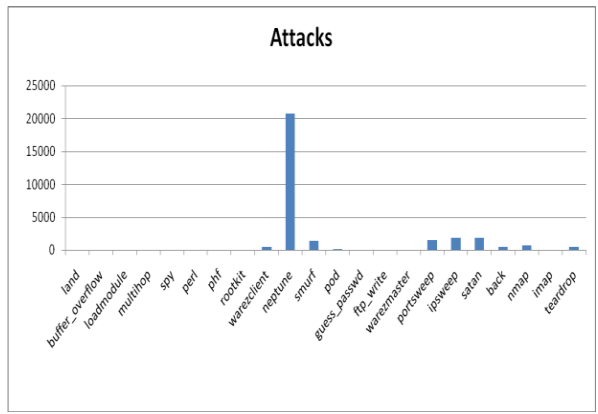

Fig. 1 Shows the distribution of the attacks in NSL-KDD dataset

## 8. Experiments and results

### 8.1. Our proposed system

We propose a hybrid feature selection algorithm based on CFS and Information gain to reduce the number of features. Our NSL-KDD dataset is first discretized using the method of Fayyad and Irani [21], then using the simple method of correlation based feature selection and a greedy search technique, a total of 10 features were selected and added to our final feature set, since CFS is not able to detect feature dependencies, information gain was used as a second step and features were ranked based on a predetermined threshold. The total number of features selected from both steps was 13. The reduced dataset was trained by a naïve Bayes classifier using the adaptive boosting technique (AdaBoost.M1) which is showed to greatly enhance the classifier performance as well as decrease the false positive rate.

We used Weka [22] as our data mining tool. Weka contains a collection of machine learning algorithms which are useful for data mining tasks like preprocessing, classification, regression, clustering, association rules, and visualization.

### 8.2. Performance Evaluation

Predictive accuracy is a poor measure and sometimes a misleading performance indicator especially in a skewed dataset [28].
Problem arises when the percentage of one class is very small compared to the other, the classifier might seem to have a high accuracy, although it fails to classify some or any of the minority class [29].

In our work we used the following two performance measures:

#### 8.2.1. F-measure

The F-measure or F-score is one of the evaluation metrics that is based on a combination of precision and recall. The larger the F-measure value, the higher the classification.

$$precision = \frac{TP}{TP+FP} \quad (4)$$

$$recall = \frac{TP}{TP+FN} \quad (5)$$

$$F_{measure} = \frac{2 \times precision \times recall}{recall+precision} \quad (6)$$

Where
- TP is the number of true positives
- FP is the number of false positives
- FN is the number of false negatives

#### 8.2.2. False Positive Rate (FPR)

False positive rate or false alarm rate is considered one of the important factors in the design of any IDs. It is the frequency of IDS reporting malicious activity when it is not.

$$FPR = \frac{FP}{FP+TN} \quad (7)$$

### 8.3. Results

All experiments were performed on a Windows platform having configuration Intel® core™ i5 CPU 2.50 GHZ, 4 GB RAM.
We used the Weka tool to evaluate our method and perform feature selection. The dataset is first discretized using the supervised discretize filter in Weka. Feature selection is done using the CFS algorithm, and a greedy search strategy is adopted which shows to select a fewer features than using the default BestFirst search strategy.

A total of 10 features are selected {4,5,7,8,10,12,30,35,36,37} and added to our final feature set. The second step involves ranking the features based on their information gain measure, and the top ranked features are selected based on a predetermined threshold.

Table 3 shows the results of applying different selection techniques on our dataset using 23 classes.

Table 3: Comparison of different selection methods using 23 classes

| Method | No. Of features | F-measure | Selected features | FPR |
|---|---|---|---|---|
| CFS+BestFirst | 18 | 97.8% | 2,3,4,5,6,7,8,10,12,23,25,29,30,35,36,37,38,40 | 0.003 |
| CFS+Greedy | 10 | 98.4% | 4,5,7,8,10,12,30,35,36,37 | 0.004 |
| Information Gain (α=0.3) | 20 | 97% | 5,3,4,30,35,29,23,34,33,6,38,25,39,26,36,12,37,24,32,2 | 0.002 |
| Gain Ratio | 19 | 97% | 8,7,4,13,26,25,39,12,30,10,38,11,2,5,29,6,27,3,35 | 0.003 |
| Correlation | 18 | 96.7% | 26,4,25,12,30,39,38,29,6,5,37,34,32,35,31,36,3,33 | 0.002 |
| CFS+IG | 13 | 98.5% | 3,4,5,7,8,10,12,23,29,30,35,36,37 | 0.006 |
| Proposed Method CFS+IG(Adaboost) | 13 | 99.3% | 3,4,5,7,8,10,12,23,29,30,35,36,37 | 0.002 |

It is clear that the total feature set is greatly reduced after feature selection method, we conducted several experiments to shows the different results obtained when using different feature selection methods for our multi-class problem.

Table 3 shows that the detection accuracy of our proposed algorithm is good but the false positive rate is high, to overcome this problem we suggested using the method of Adaptive boosting on our naïve Bayes classifier. Adaptive boosting is applied to our Naïve Bayes classifier using Weka (AdaBoost.M1) and results showed a considerable drop in the false positive rate.

In Table 4, same experiments are conducted using a 5-class dataset, which shows that our proposed hybrid feature selection algorithm delivers a higher detection rate and low false positive rate using a less number of features.

Table 4: Comparison of different selection methods using 5 classes

| Method | No. Of features | F-measure | Selected Features | FPR |
|---|---|---|---|---|
| CFS+BestFirst | 11 | 97.5% | 3,4,5,6,12,14,25,29,30,37,39 | 0.013 |
| CFS+Greedy | 11 | 97.5% | 3,4,5,6,12,14,25,29,30,37,39 | 0.013 |
| Information Gain(α=0.3) | 17 | 95% | 5,3,30,4,6,29,35,23,33,34,38,25,39,26,12,37,36 | 0.020 |
| Gain Ratio(α=0.2) | 16 | 96% | 26,25,4,12,39,30,38,6,5,29,37,11,3,22,14,35 | 0.018 |
| Correlation | 19 | 95.3% | 26,25,4,12,30,39,38,6,29,5,37,32,34,31,35,3,36,33,23 | 0.019 |
| CFS+IG | 15 | 98% | 3,4,5,6,11,18,19,23,25,26,29,30,37,38,39 | 0.041 |

To better evaluate the effectiveness of using AdaBoost, the F-measure for each attack before applying our boosting technique is listed in Table 5 which shows that attacks that belong to U2R and R2L classes are very hard to detect, while Smurf, Neptune and Back belonging to Dos class report the highest detection rate of 99%.

Table 5: shows F-measure for each attack class before applying Adaboost.M1

| *Attack* | *F-measure* | *Attack* | F-measure |
|---|---|---|---|
| Land | 0.706 | Spy | 0.0 |
| Neptune | 0.994 | Phf | 0.0 |
| Smurf | 0.999 | Warezclient | 0.892 |
| Pod | 0.977 | Guess-passwd | 0.783 |
| Back | 0.995 | Ftp-write | 0.0 |
| Teardrop | 1.0 | Warezmaster | 0.0 |
| Portsweep | 0.946 | Imap | 0.0 |
| Ipsweep | 0.953 | Buffer_overflow | 0.541 |
| Satan | 0.950 | Loadmodule | 0.0 |
| Nmap | 0.877 | Perl | 0.0 |
| Multihop | 0.0 | Rootkit | 0.0 |

Table 6: shows F-measure for each attack class after applying Adaboost.M1

| *Attack* | *F-measure* | *Attack* | F-measure |
|---|---|---|---|
| Land | 0.714 | Spy | 0.0 |
| Neptune | 0.998 | Phf | 0.8 |
| Smurf | 0.999 | Warezclient | 0.930 |
| Pod | 0.989 | Guess-passwd | 0.945 |
| Back | 0.997 | Ftp-write | 0.444 |
| Teardrop | 1.0 | Warezmaster | 0.714 |
| Portsweep | 0.982 | Imap | 0.727 |
| Ipsweep | 0.982 | Buffer-overflow | 0.667 |
| Satan | 0.970 | Loadmodule | 0.400 |
| Nmap | 0.967 | Perl | 0.0 |
| Multihop | 0.0 | Rootkit | 0.133 |

Based on the tables 5 and 6, it is clear how AdaBoost improves the false positive rates of almost all the attacks except for three attack (multihop, perl and spy) classes, boosting shows no improvement. That lies in the fact that they belong to the U2R and R2L classes which are hard to detect since they do not have any sequential patterns like DOS and Probe, they are embedded in the data packets.

## 9. Conclusions and future work

In this paper, a hybrid feature selection method of CFS and IG was used as a preprocessing step for classification. First the features are evaluated based on their correlation using the CFS method and a greedy search strategy, a total of 10 features were selected out of 41. Since CFS is not guaranteed to select all of the optimal features especially when feature dependencies exist, information gain is proposed as a second step to search for more relevant features, where features are ranked based on their relevancy and the top ranked features are selected based on a predetermined threshold. The final step involves classification using Adaptive Boosting (AdaBoost.M1) implemented in Weka and naïve Bayes as the base learner. Testing is performed using the method of 10-fold cross validation where the dataset is divided into 10 folds and each fold is used once for testing and 9 times for training. Results showed that our method delivers good detection rate and a low false positive rate when compared to other approaches addressing only 5 classes.

Our Future work is to investigate ways for improving the detection rate of the U2R and R2l attacks. We also intend to study the problem of imbalance dataset in a multi-class problem and how it affects the classification accuracy.